\documentclass[aps,prb,twocolumn]{revtex4-1}

\usepackage[dvips]{graphicx}
\usepackage[dvips]{color}
\usepackage{amsmath}
\usepackage{amssymb}
\usepackage{multirow}

\usepackage{bm}

\def\Vec#1{{\bf #1}}

\def\t#1{\textrm{#1}}
\def\n{\nonumber \\ }
\def\tensor{\otimes}
\def\Z{\mathbb{Z}}

\begin{document}

\title{Topological zero modes and Dirac points
protected by spatial symmetry and chiral symmetry}

\author{Mikito Koshino$^{1}$, 
Takahiro Morimoto$^{2}$, 
and Masatoshi Sato$^{3}$}

\affiliation{
$^1$Department of Physics, Tohoku University, 
Sendai, 980--8578, Japan
\\
$^2$Condensed Matter Theory Laboratory, RIKEN, Wako, Saitama, 351-0198,
Japan
\\
$^3$
Department of Applied Physics, Nagoya University, Nagoya 464-8603, Japan
}
\date{\today}

\begin{abstract}
We explore a new class of topologically stable zero energy modes
which are protected by coexisting chiral
and spatial symmetries.
If a chiral symmetric Hamiltonian has an additional
spatial symmetry such as reflection, inversion and rotation,
the Hamiltonian can be separated
into independent chiral-symmetric subsystems
by the eigenvalue of the space symmetry operator.
Each subsystem supports chiral zero energy modes
when a topological index assigned to the block is nonzero.
%As a result, the number of protected zero energy modes
%are generally larger than in the absence of the spatial symmetry.
By applying the argument to Bloch electron
systems, we detect band touching at
symmetric points in the Brillouin zone.
In particular, we show that Dirac nodes appearing 
in honeycomb lattice (e.g. graphene) and in
half-flux square lattice are protected by three-fold
and two-fold rotation symmetry, respectively.
We also present several examples of Dirac semimetal with isolated 
band-touching points in three-dimensional $k$-space,
which are protected by combined symmetry of rotation and reflection.
The zero mode protection by spatial symmetry
is distinct from that by the conventional winding number.
We demonstrate that
symmetry-protected band touching points emerge even though
the winding number is zero.
Finally, we identify 
relevant topological charges assigned to the gapless points.
%We show that the set of the topological numbers found here
%serves as a complete minimal set of quantum numbers 
%to label the topological distinct phases
%predicted by the algebraic classification theory.
\end{abstract}

\pacs{71.20.-b, 73.22.-f, 73.61.-r}
%71.20.-b	Electron density of states and band structure of crystalline solids
%73.22.-f	Electronic structure of nanoscale materials and related systems
%73.61.-r	Electrical properties of specific thin films

%% 
%\pacs{72.10.-d,73.20.-r,73.43.Cd}
%72.10.-d 	Theory of electronic transport; scattering mechanisms
%73.20.-r 	Electron states at surfaces and interfaces
%73.43.Cd   Theory and modeling (QHE)
%\keywords{}

\maketitle

%\section{}
%\subsection{}
%\subsubsection{}

\section{Introduction}

The chiral symmetry is 
one of the fundamental symmetries to classify the topological states 
of matter. \cite{wen89, schnyder2008classification, kitaev09} 
The symmetry relates positive and negative parts
in the energy spectrum, and a nontrivial topological nature
is linked to a singular property at zero energy.
The chiral symmetry is also called sublattice symmetry, 
because the bases are divided into
two sublattices with different eigenvalues of the chiral operator 
$\Gamma=+1$ and $-1$,
and the Hamiltonian has no matrix elements inside the same
sublattice group.
The difference between the number of 
sublattices of $\Gamma=+1$ and $-1$
is a topological index which cannot change continuously.
A nontrivial index indicates the existence of
topologically protected zero energy modes.
If the chiral Hamiltonian is defined in a phase space,
on the other hand,
we have another topological invariant defined by a winding number
(Berry phase) for a closed path.
\cite{wen89,ryu_hatsugai_2002,ryu2010topological,teo-kane10,STYY11}
Nonzero winding number is also a source of
topological objects such as
band touching points and zero energy boundary modes.

A chiral symmetric system frequently comes with other
material-dependent spatial symmetry such as reflection, rotation,
and inversion.
Recent progress in the study on 
topological phase has revealed that the existence
of the spatial symmetry enriches the
topological structure of
the system\cite{Fu11,HPB11,TZV10,HLLDBF12,SMJZ13,FHEA12,teo2013existence,ueno2013symmetry,chiu2013classification,zhang2013topological}.
The spatial symmetry often stabilizes the
band touching points which are otherwise unstable.
For example, the reflection symmetry defines topological crystalline
insulators with mirror Chern numbers \cite{HLLDBF12}, 
where an even number of stable Dirac cones exist on the
surface, \cite{xu2012observation,tanaka2012experimental,dziawa2012topological}
%as observed in SnTe materials.
which are generally unstable in 
the ordinary topological insulators.\cite{hasan-kane10,qi-zhang-rmp11}
%This contrasts with ordinary topological 
%insulators\cite{hasan-kane10,qi-zhang-rmp11}
%in which an even number of surface Dirac cones are generally unstable.
We have a similar situation in Weyl semimetals 
in three dimensions.\cite{murakami-semimetal07,Wan-semimetal11,burkov-balents11}
Weyl semimetals have stable gapless low-energy excitations that are
described by a $2\times 2$ Weyl Hamiltonian,
and the spectrum is generally gapped out when 
two Weyl nodes with opposite topological charges 
merge at the same $k$-point.
In the presence of additional spatial symmetry, however, we may have Dirac
semimetals with gapless low-energy excitations described by a 
$4\times 4$ Dirac Hamiltonian, \cite{Young-Dirac-semimetal12,Wang12,Wang13}
and it has been confirmed experimentally 
in Cd$_3$As$_2$ and Na$_3$Bi.\cite{Neupane13,Borisenko13,Liu14}
Generally, a zero energy mode or gapless mode in
a band structure is topologically stable when it is realized as an
intersect of constraints given by the secular equation in momentum
space.\cite{herring37, asano2011designing}
%In the presence of spatial symmetry, 
Sophisticated topological arguments based on the K-theory
enable us to classify possible intersects systematically as topological
obstructions, predicting gapless modes consistent with the spatial
symmetries.\cite{morimoto2013topological, shiozaki14, kobayashi14}

In this paper, we find a new class of zero energy modes
protected by the coexistence of chiral symmetry and 
spatial symmetry.
If a chiral symmetric system has an additional
symmetry such as reflection, inversion and rotation,
the Hamiltonian can be block-diagonalized into
%independent chiral-symmetric subsystems
%by the eigenvalue of the space symmetric operation.
the eigenspaces of the symmetric operation,
and each individual sector is viewed as an independent 
chiral symmetric system. 
There we can define a topological index
as the difference between the numbers of sublattices,
and a nonzero index indicates the existence of chiral zero energy modes
in that sector.
As a result, the number of total zero energy modes
are generally larger than in the absence of the spatial symmetry.

If we apply the argument to Bloch electron
systems, we can detect the existence of 
zero-energy band touching at symmetric points in the Brillouin zone.
This argument predicts the existence of gapless points
solely from the symmetry,
without even specifying the detail of the Hamiltonian.
In particular, we show that the Dirac nodes appearing 
in two-dimensional (2D) honeycomb lattice (e.g. graphene) and in
half-flux square lattice are protected by three-fold ($C_3$)
and two-fold ($C_2$) rotation symmetry, respectively.
%In a honeycomb tight-binding lattice,
%for example, a topologically protected Dirac point appears at the 
%Brillouin zone corner as a consequence of $C_3$ 
%(three-fold) rotation symmetry.
%Similarly, the Dirac cone emerging in the square lattice 
%with half flux in a plaquette
%is shown to be caused by $C_2$ rotation symmetry.
We also present examples of Dirac semimetal with isolated 
band-touching points in three-dimensional (3D) $k$-space,
which are protected by rotation and reflection symmetry.
The zero-mode protection by spatial symmetry
is distinct from that by the conventional winding number,
and we actually demonstrate in several concrete models
that symmetry-protected band touching points emerge 
even though  the winding number is zero.

%@
In the last part of the paper, we list up and classify
all independent topological invariants associated with
a given Dirac point under chiral and spatial symmetries.
They consist of winding numbers and
topological indeces (sublattice number difference)
of the subsectors of the spatial symmetry operator, 
with the redundant degrees of freedom removed.
If the spatial symmetry is of order-two
(i.e., two times of operation is proportional to identity,
like reflection and inversion),
we can use the K-theory with Clifford algebra
to identify how many quantum numbers are needed
to label all topologically distinct phases.
\cite{morimoto2013topological, shiozaki14, kobayashi14}
We explicitly show that 
a set of independent winding numbers and topological indices 
serves as complete topological charges found in the K-theory.

The paper is organized as follows.
In Sec.\ \ref{sec_gen}, we present a general formulation for
zero modes protected by
the coexistence of chiral symmetry and spatial symmetry.
We then discuss protection of the Dirac points 
in $C_3$ symmetric crystals in Sec.\ \ref{sec_c3},
that in $C_2$ symmetric crystals in Sec.\ \ref{sec_c2},
respectively.
We also argue line-node protection by additional reflection symmetry
in Sec.\ \ref{sec_ref}.
Several examples of 3D Dirac semimetal
are studied in Sec.\ \ref{sec_3d_dirac}.
In Sec.~\ref{sec:charges},
we identify
independent topological charges assigned to gapless points,
and clarify the relation to the classification theory using the Clifford
algebra.
Finally, we present a brief conclusion in Sec.\ \ref{sec_conc}.

\section{General arguments}
\label{sec_gen}

We first present a general argument
for zero modes protected by a space symmetry 
in a chiral symmetric system.
We consider a Hamiltonian $H$,
satisfying
\begin{eqnarray}
&& [H,A] =  0,
\label{eq_H_A}
\\
&& \{H,\Gamma\} =  0,
\label{eq_H_G}
\end{eqnarray}
where $\Gamma$ is the chiral operator,
and $A$ is the operator describing
the spatial symmetry of the system.
We also assume
\begin{equation}
 [\Gamma,A] = 0,
\label{eq_G_A}
\end{equation}
i.e., the sublattices belonging to $\Gamma = 1$ and $-1$
are not interchanged by $A$.

Since $[H,A]=[\Gamma,A]=0$, 
the matrices $H$, $\Gamma$ and $A$ are simultaneously block-diagonalized 
into subspaces labeled by the eigenvalues of $A$ as
\begin{align}
H &=
\begin{pmatrix}
H_{a_1} &  & & \\
 & H_{a_2} & & \\
 &  & H_{a_3}& \\
 & & & \ddots
\end{pmatrix},\n
\Gamma &=
\begin{pmatrix}
\Gamma_{a_1} &  & & \\
 & \Gamma_{a_2} & & \\
 &  & \Gamma_{a_3}& \\
 & & & \ddots
\end{pmatrix},\n
A &=
\begin{pmatrix}
a_1 &  & & \\
 & a_2 & & \\
 &  & a_3& \\
 & & & \ddots
\end{pmatrix},\
\end{align}
where $a_1,a_2,\cdots$ are the eigenvalues of $A$.
Since 
Eq.\ (\ref{eq_H_G}) requires 
$\{H_{a_i}, \Gamma_{a_i}\} =0$ for all the sectors, 
each eigenspace possesses chiral symmetry independently.
Then we can define the topological index for each sector as 
%@ chiral index -> topological index, or just 'index'
\begin{align}
\nu_{a_i}=\t{tr } \Gamma_{a_i}.
\end{align}
The index $\nu_{a_i}$ is equal to the difference of 
the chiral zero modes
of the Hamiltonian $H_{a_i}$;
\begin{align}
\nu_{a_i}&=N^+_{a_i} - N^-_{a_i},
\label{eq_nu_ai}
\end{align}
where $N^\pm_{a_i}$ are numbers of chiral zero modes satisfying
\begin{align}
H_{a_i} |u^\pm_{a_i} \rangle&=0, \n
\Gamma_{a_i} |u^\pm_{a_i} \rangle&= \pm |u^\pm_{a_i} \rangle.
\end{align}
Eq.\ (\ref{eq_nu_ai}) guarantees that
there are at least $|\nu_{a_i}|$ zero-modes in each sector,
and therefore, at least $\sum_i |\nu_{a_i}|$ zero-modes 
in the total system.

On the other hand, the topological index for the total Hamiltonian
is given by the summation over all the sub-indices as
\begin{equation}
 \nu_0 = \sum_i \nu_{a_i},
\end{equation}
which in itself guarantees $|\nu_0|$ zero-modes.
Since  $\sum_i |\nu_{a_i}| \geq |\sum_i \nu_{a_i}|$,
we can generally have more zero-modes 
in the presence of additional symmetry $A$,
than in its absence.

\begin{figure}
\begin{center}
\includegraphics[width=0.9\linewidth]{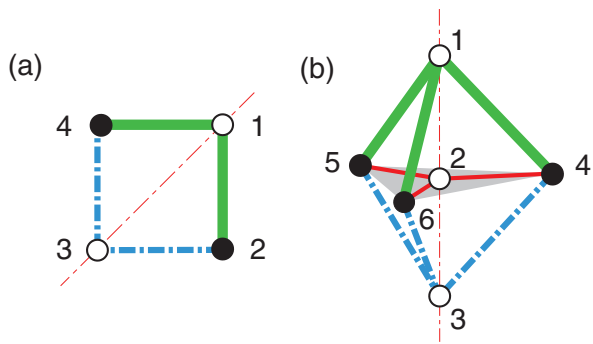}
\end{center}
\caption{
(a) 4-site model with reflection symmetry
on a diagonal axis.
(b) 6-site model with three-fold rotation symmetry.
}
\label{fig_0dim_model}
\end{figure}

% example: reflection symmetry

As the simplest example, let us consider a 4-site 
square lattice as illustrated in Fig.\ \ref{fig_0dim_model}(a).
We assume that the system is invariant under 
the reflection $R$ with respect to 
a diagonal line
connecting the site 1 and 3.
%The Hamiltonian is written as
%\begin{equation}
% H =
%\begin{pmatrix}
% & t & & t \\
%t & & t'  & \\
% & t' & & t'  \\
%t  & & t' & 
%\end{pmatrix},
%\end{equation}
We also assume that the lattice is bipartite, i.e.,
matrix elements only exist between white circles 
(sites 1 and 3) and black circles (2 and 4).
Then the system is chiral symmetric
under the chiral operator $\Gamma$ defined by
\begin{equation}
 \Gamma |i\rangle = \left\{
\begin{array}{ll}
+|i\rangle & (i=1,3)
\\ 
-|i\rangle & (i=2,4),
\end{array}
\right.
\end{equation}
where $|i\rangle$ is the state localized at the site $i$.
The operators $\Gamma$ and $R$ satisfies $[\Gamma, R] = 0$
since the black and white circles are not interchanged by $R$.
The topological index of the total system,
$\nu_0 = {\rm tr}\, \Gamma$, is obviously zero,
since there are even numbers of white and black circles.

Since $[H,R] = [\Gamma, R] = 0$,
$H$ and $\Gamma$ are block-diagonalized into subspaces
each labeled by the eigenvalues of $R$.
Each subspace is spanned by
\begin{eqnarray}
&& R={\rm even}: \quad 
|1\rangle, |3\rangle, \frac{1}{\sqrt{2}}(|2\rangle + |4\rangle)
\nonumber\\
&& R={\rm odd}: \quad
\frac{1}{\sqrt{2}}(|2\rangle - |4\rangle).
\end{eqnarray}
The topological indices of each subblock is written as
\begin{eqnarray}
&& \nu_{\rm even} = {\rm tr}\, \Gamma_{\rm even} = +1
\nonumber\\
&& \nu_{\rm odd} = {\rm tr}\, \Gamma_{\rm odd} = -1,
\end{eqnarray}
so that the number of protected zero-modes
is $|\nu_{\rm even}|+|\nu_{\rm odd}|=2$.
By breaking the reflection symmetry,
the number of zero modes is actually reduced to $|\nu_0|=0$,

%Here we should note that 
%the fixed points under the reflection 
%(i.e., the sites 1 and 3) are responsible
%to give additional zero modes.
%If all the sites are located off the symmetric axis,
%every single site has its partner at the opposite side of the axis
%and such a pair contributes 
%to a single base for each of even and odd sectors.
%As a result, we have $\nu_{\rm even} = \nu_{\rm odd}$,
%and the total number of the protected zero modes,
%$|\nu_{\rm even}|+|\nu_{\rm odd}|$
%is just equal to 
%$|\nu_{\rm tot}|=|\nu_{\rm even}+\nu_{\rm odd}|$
%which is expected in the absence of the reflection symmetry.

% example: C3 symmetry

We may consider another example having $C_3$ rotation symmetry.
Here we introduce a 6-site lattice model shown in 
Fig.\ \ref{fig_0dim_model}(b),
where the sites 1, 2 and 3 (white circles)
are located at $z$-axis,
and sites 4, 5 and 6 (black circles) are arranged in a triangle
around the origin.
The system is invariant under 
the $C_3$ rotation with respect to $z$-axis,
where the sites 1, 2, and 3 are fixed
while 4, 5, and 6 are circularly permutated.
The lattice is bipartite 
so that the Hamiltonian is chiral symmetric under $\Gamma$
defined by,
\begin{equation}
 \Gamma |i\rangle = \left\{
\begin{array}{ll}
+|i\rangle & (i=1,2,3)
\\ 
-|i\rangle & (i=4,5,6).
\end{array}
\right.
\end{equation}
Since $[H,C_3] = [\Gamma, C_3] = 0$,
$H$ and $\Gamma$ are block-diagonalized into subspaces
spanned by
\begin{eqnarray}
&& 
C_3=1: \quad
|1\rangle, |2\rangle,|3\rangle, 
\frac{1}{\sqrt{3}}(|4\rangle + |5\rangle + |6\rangle),
\nonumber\\
&& 
C_3=\omega: \quad
\frac{1}{\sqrt{3}}(|4\rangle + \omega^2 |5\rangle + \omega |6\rangle),
\nonumber\\
&& 
C_3=\omega^2: \quad
\frac{1}{\sqrt{3}}(|4\rangle + \omega |5\rangle + \omega^2 |6\rangle),
\end{eqnarray}
where $\omega = \exp(2\pi i /3)$.
The topological indices of three subspace become
\begin{eqnarray}
(\nu_{\rm 1}, \nu_{\rm \omega}, \nu_{\rm \omega^2}) 
= (2, -1, -1).
%&& \nu_{\rm 1} = {\rm tr}\, \Gamma_1 = 2
%\nonumber\\
%&& \nu_{\rm \omega} = {\rm tr}\, \Gamma_\omega = -1
%\nonumber\\
%&& \nu_{\rm \omega^2} = {\rm tr}\, \Gamma_{\omega^2} = -1.
\end{eqnarray}
The number of protected zero-modes
is $|\nu_1|+|\nu_{\omega}|+|\nu_{\omega^2}|=4$,
while we have only $|\nu_0|=0$ zero modes
in the absence of $C_3$ symmetry.
%The existence of fixed points in the symmetry operation
%is again crucial to give additional zero modes as in the previous example.
%Here the sites 1, 2, and 3 are fixed under $C_3$ rotation
%and they all contribute to the sector of $C_3=1$,
%while the sites 4, 5, and 6 are circularly interchanged
%giving a single base to each of $C_3=1,\omega,\omega^2$.
%This makes an imbalance in the topological indices among the three sectors,
%resulting in 
%non-zero values even though $\nu_1+\nu_{\omega}+\nu_{\omega^2}=0$.

\section{Dirac points in honeycomb lattice} %Dirac point protected by $C_3$ symmetry
\label{sec_c3}

%\subsection{Chiral symmetry and 
%three-fold rotation symmetry}

Now let us extend the argument in the previous section
to Bloch electron systems.
In this section, we discuss the topological protection of the Dirac points in
2D systems in the presence of three-fold rotation symmetry.
For Bloch Hamiltonian
 $H(\Vec k) = e^{-i\Vec{k} \cdot \Vec{r}} H e^{i\Vec{k} \cdot \Vec{r}}$,
the chiral symmetry and the three-fold rotation symmetry are given
by unitary operators $\Gamma$ and $C_3$ that satisfy
\begin{align}
\Gamma H(\Vec k) \Gamma^{-1}&=-H(\Vec k), \n
C_3 H(\Vec{k}) C_3^{-1} &=H(R_3(\Vec k)).
\label{eq: symmetry constraints}
\end{align}
$R_3(\Vec k)$ denotes a momentum rotated by
$120^\circ$ around the origin.
We assume the commutation relation of chiral operator and three-fold rotation,
\begin{align}
[\Gamma,C_3]=0.
\end{align}

%We can define three integer topological indices 
%that characterize gapless point of the band structure at 
%high symmetric point $\Vec k^0$.
%High symmetric point is a point in the Brillouin zone 
%that is invariant 
%under an action of three-fold rotation; $R_3(\Vec k^0)=\Vec k^0$.
%We assume gapped band structure away from $\Vec k^0$.

%%%

Let us consider the high symmetric point in the Brillouin zone 
that is invariant 
under an action of three-fold rotation; $R_3(\Vec k^0)=\Vec k^0$.
There Eq.~(\ref{eq: symmetry constraints}) reduces to
\begin{align}
\{H(\Vec k^0), \Gamma\} = 0, \n
[H(\Vec k^0), C_3] = 0,
\end{align}
and we can apply the previous argument to $H = H(\Vec{k}^0)$.
%Since $[H,C_3]=[\Gamma,C_3]=0$, 
We simultaneously block-diagonalize 
$H$, $\Gamma$ and $C_3$  
into three sectors each labeled by an eigenvalue of $C_3$,
and define a topological index for each eigenspace as
\begin{align}
\nu_a=\t{tr } \Gamma_a,
\end{align}
with $a=1,\omega,\omega^2$.
If $\sum_a |\nu_a|$ is non-zero,
it requires an existence of chiral zero modes of $H(\Vec k^0)$,
i.e., we have a topologically stable gap closing at $\Vec k^0$
protected by the chiral symmetry and $C_3$ symmetry.

%\begin{align}
%H(\Vec k^0)&=
%\begin{pmatrix}
%H_1 & 0 & 0 \\
%0 & H_\omega & 0 \\
%%0 & 0 & H_{\omega^2}
%\end{pmatrix}, \n
%\Gamma&=
%\begin{pmatrix}
%\Gamma_1 & 0 & 0 \\
%0 & \Gamma_\omega & 0 \\
%0 & 0 & \Gamma_{\omega^2}
%\end{pmatrix}, \n
%C_3&=
%\begin{pmatrix}
%1 & 0 & 0 \\
%0 & \omega & 0 \\
%0 & 0 & \omega^2
%\end{pmatrix}.
%\end{align}

\begin{figure}
\begin{center}
\includegraphics[width=0.55\linewidth]{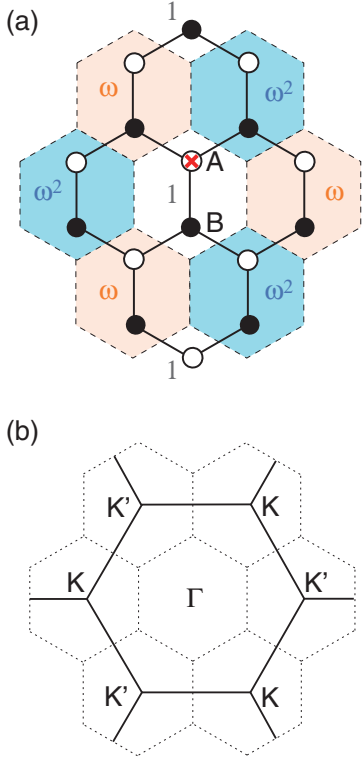}
\end{center}
\caption{
(a) Honeycomb lattice with
the nearest neighbor hopping.
Unit cell is indicated by dashed hexagon,
and shading represents the Bloch phase factor
$\exp(i\Vec{K}\cdot \Vec{r})=1,\omega,\omega^2$ 
at $K$ point.
(b) Brillouin zone for the honeycomb lattice.
Dotted, small hexagon is the reduced Brillouin zone
corresponding to $\sqrt{3}\times\sqrt{3}$ superlattice
in Figs.\ \ref{fig_honeycomb_superlattice}(a) and (b).
}
\label{fig_graphene}
\end{figure}

%\subsection{Graphene with Kekule distortion}

Graphene is the simplest example of 
the band touching protected by $C_3$ symmetry.
Let us consider a tight-binding honeycomb lattice
with the nearest neighbor hopping as shown in Fig.\ \ref{fig_graphene}(a).
The unit cell is composed of non-equivalent A and B sublattices.
The Hamiltonian is chiral symmetric in that A is only connected to B,
and the system is obviously invariant under three-fold rotation $C_3$.
$C_3$ commutes with $\Gamma$ because the rotation does not
interchange A and B sublattices.
The Brillouin zone corners $K$ and $K'$,
shown in Fig.\ \ref{fig_graphene}(b), 
are fixed 
in $C_3$ rotation so that we can apply the above argument to 
these points.
We define the Bloch wave basis as
\begin{eqnarray}
&& |X\rangle = 
\frac{1}{\sqrt{N}}\sum_{\Vec{R}_{X}} e^{i\Vec{k}\cdot\Vec{R}_{X}}
|\Vec{R}_{X}\rangle \quad  (X=A,B),
\end{eqnarray}
where $\Vec{k}$ is the Bloch wave vector ($K$ or $K'$),
$|\Vec{R}_{X}\rangle$ is the atomic state at the position $\Vec{R}_{X}$,
and $N$ is the number of unit cells in the whole system.
In the basis of $\{|A\rangle, |B\rangle\}$,
the chiral operator is written as
\begin{align}
\Gamma =
\begin{pmatrix}
1 & \\
& -1 \\
\end{pmatrix}.
\end{align}
If we set the rotation center at $A$ site,
the rotation $C_3$ is written as
\begin{eqnarray}
&&C_3 =
\begin{pmatrix}
1 & \\
& \omega  
% & \omega^2  ! fixed on 140609 
\end{pmatrix}
\t{for } K, 
\quad
\begin{pmatrix}
1 & \\
& \omega^2 
% & \omega   ! fixed on 140609 
\end{pmatrix}
\t{for } K'.
\end{eqnarray}
This is actually derived by 
considering the change of the  Bloch factor
in the rotation [Fig.\ \ref{fig_graphene}(a) for $K$ point].
Therefore, the topological indices are obtained as 
\begin{align}
(\nu_1,\nu_\omega,\nu_{\omega^2}) &=
\begin{cases}
(1,-1,0) & \t{for } K, \\
(1,0,-1) & \t{for } K'.
%(1,0,-1) & \t{for } K, \\
%(1,-1,0) & \t{for } K'.
\end{cases}
\end{align}
This requires 
 two zero-modes at each of $K$ and $K'$, 
which are nothing but the gapless Dirac nodes.
\cite{mcclure1956diamagnetism,JPSJ.74.777}
Note that the band touching is deduced purely from the symmetry
in the Bloch bases,
without specifying detailed Hamiltonian matrix.

\begin{figure}
\begin{center}
\includegraphics[width=0.6\linewidth]{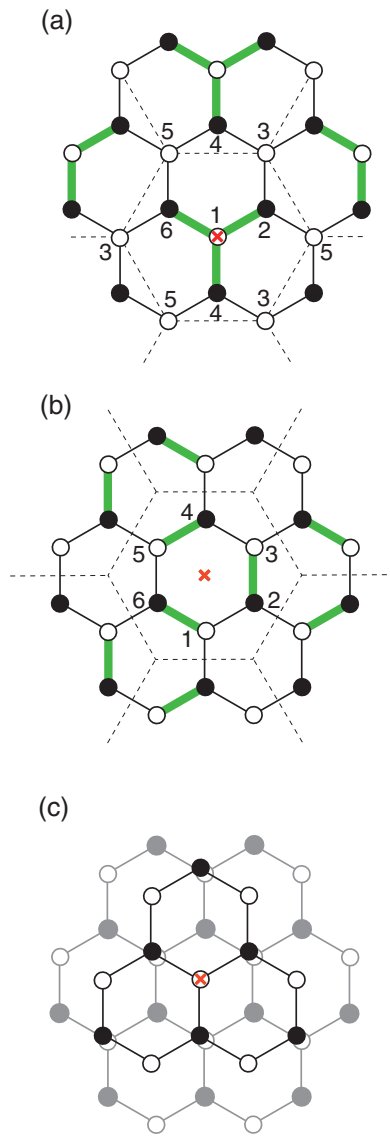}
\end{center}
\caption{
(a, b) $\sqrt{3}\times\sqrt{3}$ superlattice unit
cell of graphene,
with possible lattice distortion under 
(a) $C_3$ and (b) $C'_3$ symmetry.
The center of the rotation is indicated by a cross mark.
(c) Structure of graphite,
where black and gray layers stack alternatively.
}
\label{fig_honeycomb_superlattice}
\end{figure}

In this particular case, 
the gaplessness at  $K$ and $K'$ can also be concluded from the 
non-trivial winding number $\nu_W = \pm 1$
around $K$ and $K'$, respectively.
[For details of the winding number, see Sec.~\ref{sec:charges}A]
%However, the two different arguments 
%(winding number and the spatial symmetry)
%do not always lead to the same conclusion
%because $\nu_0$ is not equivalent with 
%$\nu_1,\nu_\omega,\nu_{\omega^2}$.
However, these two different arguments are not generally equivalent,
and actually $C_3$-protected band touching may occur even though
the winding number is zero, as shown in the following.
Let us consider a tight-binding honeycomb lattice
with $\sqrt{3}\times\sqrt{3}$ superlattice distortion
as shown in Figs.\ \ref{fig_honeycomb_superlattice} (a) and (b),
where the hopping amplitudes for thin and thick bonds 
are differentiated.
In accordance with the enlarged unit cell,
the Brillouin zone is folded as 
shown in Fig.\ \ref{fig_graphene} (b),
where the original $K$, $K'$ and $\Gamma$
points are folded onto the new $\Gamma$-point.
Then $\nu_W$ around the $\Gamma$-point is contributed from
$\pm 1$ around the original $K$ and $K'$, respectively, 
so that we have trivial winding number $\nu_W=0$ as a whole.
%Therefore we have no $\nu_0$-protection of the Dirac point.
However, we can show that the Dirac point remains ungapped
even in the presence of the superlattice distortion,
when the system keeps a certain three-fold rotation symmetry.
We consider two different types of rotations.
\begin{align} 
& C_3: \, \mbox{120$^\circ$ rotation around $A$ site.}
\n
& C'_3: \, \mbox{120$^\circ$ rotation around the center of hexagon.}
\nonumber
\end{align}
Figs.\ \ref{fig_honeycomb_superlattice}(a) and \ref{fig_honeycomb_superlattice}(b)
are examples of the lattice distortion under
$C_3$ and $C'_3$ symmetry, respectively.
The latter case, Fig.\ \ref{fig_honeycomb_superlattice}(b),
is so-called Kekul\'{e} distortion.

The unit cell contains six atoms as depicted
in Fig.~\ref{fig_honeycomb_superlattice}.
In the basis of
$\{ |1\rangle,|2\rangle,\cdots ,|6\rangle\}$,
the chiral operator is given by
\begin{align}
\Gamma&=
\begin{pmatrix}
1 &&&&& \\
& -1 &&&& \\
&& 1 &&& \\
&&&-1 && \\
&&&& 1 & \\
&&&&&-1  \\
\end{pmatrix},
\end{align}
and the three-fold rotation at $\Gamma$-point is
represented by
\begin{align}
C_3&=
\begin{pmatrix}
1 &&&&& \\
&  &&1&& \\
&& 1 &&& \\
&&& &&1 \\
&&&& 1 & \\
&1&&&&  \\
\end{pmatrix}, \n
C'_3&=
\begin{pmatrix}
 &&&&1& \\
& &&&&1 \\
1&& &&& \\
&1&& && \\
&&1&& & \\
&&&1&&  \\
\end{pmatrix}.
\end{align}
The topological indices of three sectors are given by
\begin{align}
(\nu_1,\nu_\omega,\nu_{\omega^2}) &=
\begin{cases}
(2,-1,-1) & \t{for } C_3, \\
(0,0,0) & \t{for } C'_3.
\label{eq_3nus}
\end{cases}
\end{align}
Non-trivial topological indices in $C_3$ symmetry
requires four zero-modes, indicating that
the two Dirac points are protected.
%Note that the trivial winding number $\nu_0 = 0$
%is not the origin of the protection of the gapless point.
In $C'_3$ symmetry, on the other hand,
the topological indices are all zero and the energy band is gapped out.
The situation of $C_3$ symmetry 
closely resembles the 6-site model in the previous section,
where the fixed points in the rotation (the sites 1, 3 and 5)
all contribute to the sector of $C_3=1$,
leading to an imbalance in the topological indices among the three sectors.
In contrast, all the sites are circularly interchanged 
in $C'_3$ rotation,
resulting in $\nu_1=\nu_\omega=\nu_{\omega^2}=\nu_0/3=0$.

We can derive the same conclusion alternatively by starting from
$4\times 4$ low-energy effective Hamiltonian,
\begin{align}
H=k_x \sigma_x \tau_z + k_y \sigma_y,
\label{eq: Dirac H for graphene}
\end{align}
where Pauli matrices $\sigma$ and $\tau$ 
span the sublattice ($A, B$) and the valley $(K, K')$ 
degrees of freedom, respectively.
The dimension of the matrix is smaller than the previous argument
($6\times6$)
because we exclude the two high-energy bases from the original $\Gamma$-point,
which do not contribute to the topological indices.
The chiral operator is given by
\begin{align}
\Gamma =\sigma_z =
\begin{pmatrix}
1 &&& \\
& -1 && \\
&& 1 & \\
&&&-1 \\
\end{pmatrix}.
\end{align}
The matrices for $C_3$ and $C'_3$ are 
derived by considering the Bloch factor in the original lattice
model as,
\begin{eqnarray}
&&C_3 =
\begin{pmatrix}
1 &&& \\
& \omega && \\   % & \omega^2 && \\ 
&& 1 & \\
&&& \omega^2 \\   % & \omega && \\  
\end{pmatrix}
=\exp\left[-\frac{\pi i}{3}(\sigma_z-1)\tau_z\right].
% =\exp\left[\frac{\pi i}{3}(\sigma_z-1)\tau_z\right].
\\
&&C'_3
=
\begin{pmatrix}
\omega &&& \\
& \omega^2 && \\
&& \omega^2 & \\
&&& \omega \\
%\omega^2 &&& \\
%& \omega && \\
%&& \omega & \\
%&&& \omega^2 \\
\end{pmatrix}
=\exp\left[\frac{2\pi i}{3}\sigma_z\tau_z\right],
%=\exp\left[-\frac{2\pi i}{3}\sigma_z\tau_z\right],
\end{eqnarray}
The topological indices are immediately shown to 
be equivalent to Eq.\ (\ref{eq_3nus}).

Possible mass terms under the chiral symmetry
which gap out the Hamiltonian of Eq.\ (\ref{eq: Dirac H for graphene})
anti-commute with both $H$ and $\Gamma$.
In the present case we have two such terms,
\begin{align}
\delta H= m_x \sigma_x \tau_x + m_y \sigma_x \tau_y.
\label{eq:massterm}
\end{align}
Since $\delta H$ commutes with $C_3'$ but not with $C_3$,
these terms can exist only in $C'_3$ symmetry.
This exactly corresponds to the fact that
the Dirac point is not protected in $C'_3$ symmetry.
Actually, Kekul\'{e} distortions depicted in Fig.~\ref{fig_honeycomb_superlattice}(b)
give rise to mass terms $\delta H$ 
and gap out the Dirac point.

The argument can be directly extended to
a 3D crystal with $C_3$ symmetry.
There the band touching point forms a line node
on $C_3$ symmetric axis
in 3D Brillouin zone.
A typical example of this is a bulk graphite,
where graphene layers are stacked in an alternative way 
between black and gray layers as in Fig.\ \ref{fig_honeycomb_superlattice}(c).
%We do not consider the lattice distortion
%and take the original in-plane unit cell.
When we consider the three-fold rotation symmetry
around the center of hexagon of a gray layer
(cross mark in the figure),
the topological indices are obtained as 
\begin{align}
(\nu_1,\nu_\omega,\nu_{\omega^2}) &=
\begin{cases}
(1,0,-1) & \t{for } K, \\
(1,-1,0) & \t{for } K'.
%(1,-1,0) & \t{for } K, \\
%(1,0,-1) & \t{for } K',
\end{cases}
\end{align}
which guarantees two line nodes  at $K$ and $K'$ parallel to $k_z$ direction.
A real graphite is not exactly chiral symmetric 
because of some minor hopping amplitudes
between black and black (white and white) atoms.
As a result, the line node slightly disperses in $k_z$ axis
giving electron and hole pockets at zero energy.\cite{dresselhaus2002intercalation}

\begin{figure}
\begin{center}
\includegraphics[width=0.65\linewidth]{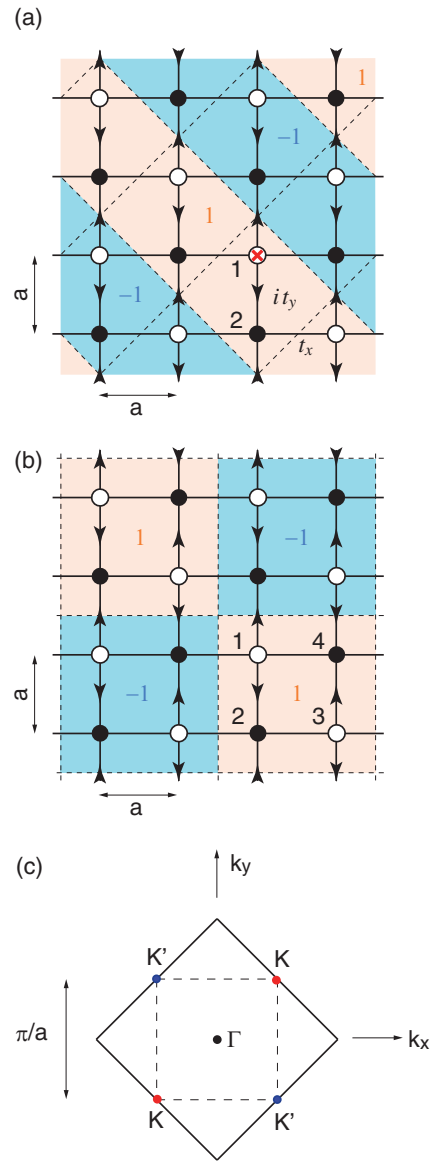}
\end{center}
\caption{
(a) Square lattice with a half magnetic flux penetrating unit cell.
Unit cell is indicated by a dashed diamond,
and shading represents the Bloch phase factor
for $K$ point.
(b) The same system with double unit cell.
(c) Brillouin zone for the single unit cell in (a).
Dashed square is the reduced Brillouin zone corresponding to
the double unit cell in (b).
}
\label{fig_half_flux}
\end{figure}

%%%%%%%%%%%%%%%%%%%%%%%%%%
\section{Dirac points in half-flux square lattice}
\label{sec_c2}

The square lattice with half magnetic flux penetrating a unit cell
is another well-known example having gapless Dirac nodes.\cite{affleck1988large}
The band touching in this system can also be explained 
by a similar argument, 
in terms of the chiral symmetry and $C_2$ rotation symmetry.
We consider a lattice Hamiltonian illustrated 
in Fig.\ \ref{fig_half_flux}(a).
The unit cell is represented by a dashed diamond
including site 1 and 2 inside.
The hopping integral along horizontal bond is all identical to $t_x$,
while the vertical hopping depends on the direction,
and is given by $i t_y$ for the hopping in the direction of the arrow.
An electron always acquire the factor $-1$ when moving around 
any single plaquette, so that it is equivalent to half magnetic flux
penetrating a unit cell.

The system is $C_2$-rotation symmetric with respect to
an arbitrary atomic site,
and the rotation commutes with the chiral operator
since it does not interchange the sublattices.
In the reciprocal space [Fig.\ \ref{fig_half_flux}(c)],
the points $K:(\pi/(2a),\pi/(2a))$ and 
$K':(-\pi/(2a),\pi/(2a))$ are both invariant in the $C_2$ rotation,
and we apply the general argument to these points.
In the basis of $\{|1\rangle, |2\rangle\}$,
the chiral operator is written as
\begin{align}
\Gamma =
\begin{pmatrix}
1 & \\
& -1 \\
\end{pmatrix},
\end{align}
and the $C_2$ rotation with respect to site 1 is
\begin{eqnarray}
&&C_2 =
\begin{pmatrix}
1 & \\
& -1
\end{pmatrix}  \quad \t{for $K$ and $K'$}.
\end{eqnarray}
Thus the topological indices are
\begin{align}
(\nu_{\rm even},\nu_{\rm odd}) = (1,-1) \quad \t{for $K$ and $K'$}, 
\end{align}
where even and odd specify the eigenvalue of $C_2$ rotation
$+1$ and $-1$, respectively.
As a result, we have two zero-modes at each of $K$ and $K'$, 
corresponding to the band touching points.

Similarly to the honeycomb lattice in the previous section, 
we may consider the stability of the Dirac points
under possible lattice distortions
for the double unit cell shown in  Fig.\ \ref{fig_half_flux}(b).
In the reciprocal space, $K$ and $K'$ merge at the same corner point
and the total winding number becomes zero.
We consider two types of rotations.
\begin{align} 
& C_2: \, \mbox{180$^\circ$ rotation around site 1.}
\n
& C'_2: \, \mbox{180$^\circ$ rotation around the center of square.}
\nonumber
\end{align}
In the basis of
$\{ |1\rangle,|2\rangle,|3\rangle\,|4\rangle\}$,
the chiral operator is given by
\begin{align}
\Gamma&=
\begin{pmatrix}
1 &&& \\
& -1 && \\
&& 1 & \\
&&&-1  \\
\end{pmatrix},
\end{align}
and the 180$^\circ$ rotation at the merged $k$-point is
represented by
\begin{align}
C_2 &=
\begin{pmatrix}
1 &&& \\
& -1 && \\
&&  1 & \\
&&&  -1 \\
\end{pmatrix}, \n
C'_2 &=
\begin{pmatrix}
&&1& \\
& &&1 \\
1&&  & \\
&1&&   \\
\end{pmatrix}.
\end{align}
The topological indices are given by
\begin{align}
(\nu_{\rm even},\nu_{\rm odd}) &=
\begin{cases}
(2,-2) & \t{for } C_2, \\
(0,0) & \t{for } C'_2,
\label{eq_2nus}
\end{cases}
\end{align}
so that the two Dirac points are protected
in $C_2$ symmetry,
while not in $C'_2$ symmetry.

The same conclusion can be reproduced in terms of the low energy
effective Hamiltonian, in a manner similar to Sec.~\ref{sec_c3}.
The effective Hamiltonian and the chiral symmetry are given by 
\begin{eqnarray}
H=k_x\sigma_x\tau_z +k_y\sigma_y,
\quad
\Gamma=\sigma_z
\end{eqnarray}
with $\sigma$ and $\tau$ spanning the sublattice $(|1\rangle,
|2\rangle)$ and the valley $(K,K')$ of the $\pi$-flux lattice.
By taking into account the Bloch factor properly, the two-fold rotations
$C_2$, $C_2'$ are identified as 
\begin{eqnarray}
C_2=\sigma_z,
\quad
C_2'=\sigma_z\tau_z. 
\end{eqnarray}
Since the effective Hamiltonian and the chiral symmetry take the same
forms as those in the honeycomb lattice case,  possible mass terms consistent
with the chiral symmetry are given by the same Eq.(\ref{eq:massterm}).
Those mass terms are apparently inconsistent with  the $C_2$ symmetry
above, but consistent with the $C_2'$ symmetry.
Thus, between the two types of rotations, only the $C_2$ symmetry does not
allow these mass terms, keeping the Dirac points gapless.  

\section{Line node protected by reflection symmetry}
\label{sec_ref}

\begin{figure}
\begin{center}
\includegraphics[width=0.5\linewidth]{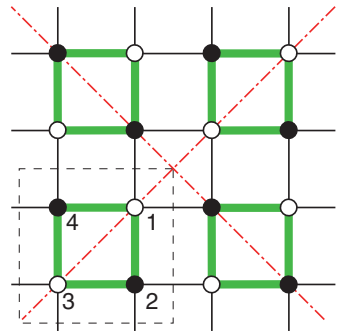}
\end{center}
\caption{
Square lattice model with the reflection symmetry.
Dashed square indicates a unit cell and the diagonal line
is a symmetry axis.
}
\label{fig_square_lattice}
\end{figure}

As another example,
we consider a 2D lattice
with the reflection symmetry.
In this case, the band touching is protected 
on the diagonal lines in 2D Brillouin zone,
and form line nodes.
%Let us Bloch Hamiltonian $H(\Vec k)$
%that satisfies
%\begin{align}
%\Gamma H(\Vec k) \Gamma^{-1}&=-H(\Vec k), \n
%R H(R(\Vec k)) R^{-1} &=H(\Vec k),
%\end{align}
%where $\Gamma$ is the chiral operator,
%and $R$ is the reflection on a certain symmetric axis.
%We assume a commutation relation of the chiral operator and the reflection,
%\begin{align}
%[\Gamma,R]=0.
%\end{align}
%
%On the reflection-symmetric point $R(\Vec k^0)=\Vec k^0$,
%we have
%\begin{align}
%\{H(\Vec k^0), \Gamma\} = 0, \n
%[H(\Vec k^0), R] = 0.
%\end{align}
%We can then simultaneously block-diagonalize 
%$H(\Vec{k}^0)$, $\Gamma$ and $R$  
%into even and odd sectors with respect to the eigenvalue of $R$,
%and define a topological indices $\nu_{\rm even}, \nu_{\rm odd}$.
%If $|\nu_{\rm even}|+ |\nu_{\rm odd}|$ is non-zero,
%we have a band gap closing at $\Vec k^0$.
We take a lattice model
as illustrated in Fig.\ \ref{fig_square_lattice},
where the unit cell is composed of four sublattices
from 1 to 4, and the structure is reflection symmetric
with respect to the diagonal lines. 
In the basis of
$\{ |1\rangle,|2\rangle,|3\rangle,|4\rangle\}$,
the chiral operator is given by
\begin{align}
\Gamma&=
\begin{pmatrix}
1 &&& \\
& -1 && \\
&& 1 & \\
&&&-1  \\
\end{pmatrix}.
\end{align}
We consider the reflection $R$ with respect to
the line connecting the sites 1 and 3.
The fixed $k$-points under $R$ are given by $\Vec{k}_0 = (k,k)$
with arbitrary $k$.
There the matrix for $R$ is written as
\begin{align}
R&=
\begin{pmatrix}
1 &&& \\
&  &&1 \\
&& 1 & \\
&1 &&  \\
\end{pmatrix}.
\end{align}
The situation is exactly the same as the four-site model
in Sec.\ \ref{sec_gen}, and
the topological indices of two sectors become
\begin{align}
(\nu_{\rm even},\nu_{\rm odd}) &= (1,-1).
\end{align}
Since $|\nu_{\rm even}|+|\nu_{\rm odd}|=2$,
two energy bands are touching 
along the diagonal axis in the Brillouin zone.
The same argument applies to the reflection for another
diagonal line, giving a line node at $(k,-k)$.

%Since a winding number $\nu_0 = \nu_{\rm even} + \nu_{\rm odd}$ is 0,
%the band touching is a consequence of 
%coexistence of the reflection symmetry and the chiral symmetry.

\begin{figure}
\begin{center}
\includegraphics[width=0.95\linewidth]{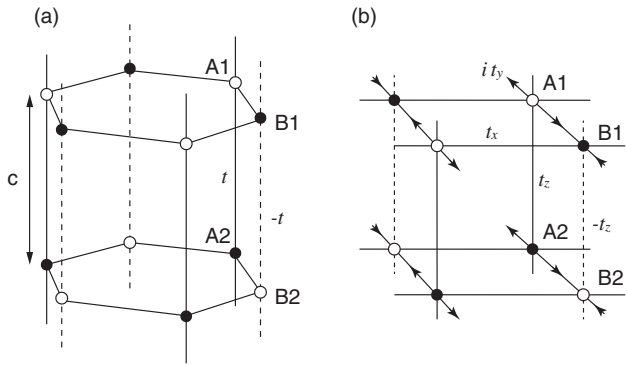}
\end{center}
\caption{
(a) Stacked honeycomb lattices 
with staggered interlayer coupling.
(b) Cubic lattice 
with a half magnetic flux penetrating every square plaquette.
}
\label{fig_3d_dirac}
\end{figure}

\section{Dirac points in three dimensions}
\label{sec_3d_dirac}

Here we present some examples %@@
of 3D Dirac system, 
where the band touching occurs at isolated $k$-points 
in 3D Brillouin zone.
First we consider a stack of 
honeycomb lattices with staggered interlayer coupling
as illustrated in Fig.\ \ref{fig_3d_dirac}(a).
Here the honeycomb layers are vertically stacked
at interlayer spacing $c$,
and the vertical hopping between the neighboring layers 
is given by $t$ and $-t$ for $A$ and $B$ sublattices,
respectively. 
The smallest unit cell of this system is given by
$A$ and $B$ on a single layer,
while we here take a double unit cell
including $A1,B1,A2,B2$,
so that the Hamiltonian becomes chiral symmetric
by grouping $(A1,B2)$ into $\Gamma=+1$,
and $(B1,A2)$ into $-1$.
%At each of $K$ and $K'$ valleys, 
The effective Hamiltonian is given by
\begin{equation}
H = k_x \sigma_x\tau_z + k_y\sigma_y+ 2 t \cos(k_z c)\sigma_z \rho_x, 
\label{eq_H_3d_dirac}
\end{equation}
where Pauli matrices $\sigma$ and $\rho$ 
span the sublattice ($A, B$) and the layer $(1, 2)$ 
degrees of freedom, respectively,
and $\tau_z=\pm 1$ is the valley indices for $K$ and $K'$,
respectively.
Equation \ (\ref{eq_H_3d_dirac}) has a gapless node at 
$\Vec{k}^0=(0,0,\pi/(2c))$, and two Dirac cones are degenerate 
at this point.
Note that the lattice period in $z$ direction is $2c$,
so that $-\Vec{k}^0$ is equivalent to $\Vec{k}^0$.

The gapless point at $\Vec{k}^0$
can be concluded from the symmetry 
argument without the band calculation.
The chiral operator is given by $\Gamma = \rho_z\sigma_z$,
which obviously anticommutes with the Hamiltonian.
We consider $C_3$ rotation with respect to $A1$-$A2$ axis,
and also the reflection $R_z$ with respect to $A1$-$B1$ layer.
The Hamiltonian is invariant 
and also the point $\Vec{k}^0$ is fixed
under these operations.
Now we consider a combined operation $C_3 R_z$
at $\Vec{k}_0$.
Since  $(C_3 R_z)^6 = 1$, the eigenvalues of $C_3 R_z$
can be either of $\pm 1, \pm\omega, \pm\omega^2$.
For $K$-valley, for example, the matrix of
 $C_3 R_z$ in a basis of $\{
 |A1\rangle,|B1\rangle,|A2\rangle,|B2\rangle\}$ becomes
\begin{equation}
C_3 R_z = 
{\rm diag}(1,\omega,-1,-\omega),
%{\rm diag}(1,\omega^2,-1,-\omega^2),
\label{eq: charges C3Rz}
\end{equation}
i.e., the four sublattices are classified to all different sectors.
The number of zero modes is $\sum_a |\nu_a| = 4$, 
which guarantees the existence of doubly degenerate Dirac nodes.
The argument equally applies to more general cases
where the vertical hopping at $A$ and $B$-sites
are given by $t_A$ and $t_B$ (instead of $t$ and $-t$),
respectively.
%although the central Dirac cone generally overlaps 
%with other part of the energy band 
%when $t_A$ and $t_B$ have the same sign.

%%%%%%%%%%%@@

We can create another example of 3D Dirac nodes
by stacking 2D $\pi$-flux lattice
in Sec.\ \ref{sec_c2} 
with staggered interlayer coupling.
The model is illustrated in Fig.\ \ref{fig_3d_dirac}(b),
where $\pi$-flux lattices are vertically stacked
with the hopping $t_z$ and $-t_z$ for $A$ and $B$ sublattices,
respectively. 
The system can be viewed as a cubic lattice
with a half magnetic flux threading every single 
square plaquette.
We take a unit cell composed of $A1,B1,A2,B2$,
and group $(A1,B2)$ into $\Gamma=+1$, and $(B1,A2)$ into $-1$
so that the Hamiltonian becomes chiral symmetric.
We have band touching at $K:\pi/(2a)(1,1,1)$ and
$K':\pi/(2a)(-1,1,1)$,
and the effective Hamiltonian near these point nodes is given by
\begin{equation}
H = k_x \sigma_x\tau_z + k_y\sigma_y - k_z \sigma_z \rho_x, 
\label{eq_H_3d_pi-flux}
\end{equation}
where Pauli matrices $\sigma$ and $\rho$ 
span the sublattice ($A, B$) and the layer $(1, 2)$ 
degrees of freedom, respectively,
and $\tau_z=\pm 1$ is the valley indices for $K$ and $K'$,
respectively.
The chiral operator is given by $\Gamma = \rho_z\sigma_z$.

The gapless point in this model is protected by the inversion symmetry
$P = C_2 R_z$.
If we consider the inversion $P$ with respect to $A1$ site,
$K$ and $K'$ are both invariant,
and we can write $P = \rho_z \sigma_z$ at these points. 
We then find $(\nu_{\rm even},\nu_{\rm odd})=(2,-2)$,
and thus we have doubly degenerate Dirac nodes at each of $K$ and $K'$.

%%%%%%%%%%%%%%%%%%%%%%%%

\section{Classification of topological charges \label{sec:charges}}

In this section, we present general arguments to classify the 
Dirac points in the
presence of chiral symmetry and spatial symmetry. We identify relevant
topological numbers associated with protection
of the Dirac points.
In Table \ref{table: topological charges},
we summarize our results on topological charges 
of the Dirac points obtained in this section.

\begin{table}[tb]
\begin{center}
\caption{\label{table: topological charges}
Topological charges of the Dirac points in the presence of
chiral symmetry and spatial symmetry.
We assume that symmetry operators commute with each other.
}
\begin{tabular}[t]{ccc}
\hline \hline
~Dimensions~ & ~Symmetries~ & ~Charges~ \\
\hline
2D & $\Gamma, C_N$ & $\mathbb{Z}^N$ \\
3D & $\Gamma, C_3 R_z$ & $\mathbb{Z}^3$ \\
3D & $\Gamma, C_2, R_z$ & $\mathbb{Z}^2$ \\
3D & $\Gamma, P$ & $\mathbb{Z}$ \\
\hline \hline
\end{tabular}
\end{center}
\end{table}

\subsection{Class AIII+$C_N$ in 2D}
First, let us study 2D Dirac points 
in class AIII systems (possessing chiral symmetry $\Gamma$)
with additional $N$-fold rotation symmetry $C_N$.
We assume the commutation relation $[\Gamma,C_N]=0$.
the Dirac points in Sec.~\ref{sec_c3} and Sec.~\ref{sec_c2} are of this class.

In the presence of the chiral symmetry, we can define a winding number
for a circle $S^1$ surrounding 
the Dirac point in the Brillouin zone.\cite{wen89}
When the circle $S^1$ is parameterized by $\theta
\in [0 , 2\pi)$, the winding number is given by
\begin{eqnarray}
\nu_W=\frac{1}{4\pi i}\oint_{S^1}d\theta{\rm tr}
\left[
\Gamma H^{-1}(\Vec{k}(\theta))
\partial_\theta H(\Vec{k}(\theta))
\right].  
\label{eq: winding number}
\end{eqnarray}
Here the Hamiltonian is gapped on $S^1$ so the inverse $H^{-1}(\Vec{k}(\theta))$ is well-defined. 
%
%For a circle $S^1$ surrounding the Dirac point in the Brillouin zone,
%we can define a winding number $n_W$ in class AIII as
%follows.\cite{schnyder2008classification, ryu2010topological} 
In a basis where the chiral operator $\Gamma$ is diagonal, 
\begin{align}
\Gamma=
\begin{pmatrix}
1 & 0 \\
0 & -1 \\
\end{pmatrix},
\end{align} 
the Hamiltonian takes an off-diagonal form written as
\begin{align}
H(\Vec k)=
\begin{pmatrix}
0 & D^{\dagger}(\Vec k) \\
D(\Vec k) & 0 \\
\end{pmatrix}.
\end{align}
Here $\t{tr } \Gamma$ must be zero (i.e., $D$ is a square matrix),
since otherwise zero energy states
remain independently of $\Vec{k}$.
%and the spectrum is not gapped.  %@@@
The winding number is then recast into
\begin{eqnarray}
\nu_W=\frac{1}{2\pi}{\rm Im}\left[
\oint_{S^1}
d\theta \partial_{\theta}\ln
{\rm det}D(\Vec{k}(\theta))
\right].
\end{eqnarray}
It is evident that $\nu_W$ is quantized
to an integer since the phase change of ${\rm det}\,D(\Vec{k}(\theta))$
around $S^1$ must be a multiple of $2\pi$.  

%with a Dirac operator $D$.
%If we take $\Vec k$ on the circle $S$,
%the Hamiltonian is gapped and 
%$D$ is a full rank matrix.
%Then we have a singular value decomposition of $D$ as
%\begin{align}
%D=U\Sigma V^\dagger,
%\end{align}
%with unitary matrices $U$ and $V$,
%and a diagonal matrix $\Sigma$ whose diagonal entries are
%real and positive. 
%Actually diagonal entries of $\Sigma$ are
%positive eigenvalues of $H$.
%Now we define a flattened Hamiltonian
%\begin{align}
%\widetilde H(\Vec k) &=
%\begin{pmatrix}
%0 & q(\Vec k) \\
%q^\dagger (\Vec k) & 0\\
%\end{pmatrix}, \n
%q(\Vec k) &=
%U(\Vec k) V^\dagger (\Vec k),
%\end{align}
%in which positive and negative eigenvalues are flattened to $+1$ or $-1$,
%with preserving wavefunctions.
%When the circle $S^1$ around the Dirac point is parameterized by $\theta \in [0 , 2\pi)$,
%we have a winding number $\nu_W$ defined for $q[\Vec k(\theta)]$ as
%\begin{align}
%\nu_W = 
%\frac{1}{2\pi i}
%\oint_S d\theta q[\Vec k(\theta)]^\dagger \partial_\theta 
%q[\Vec k(\theta)].
%
%\end{align}
%This corresponds to Berry phase around Dirac point at $\Vec k^0$
%in a unit of $\pi$.
%Berry phase is quantized in 2D system with chiral symmetry
%and its non-triviality indicates a stable gapless point within the circle.

As we have seen in Sec.~\ref{sec_gen},
by making use of rotation symmetry $C_N$, 
we can further define topological indices 
$\nu_{a_n}\, [a_n = \exp(2\pi n i / N),\, n=0,1,2,\cdots,N-1]$ 
by Eq.~(\ref{eq_nu_ai}),
for the Dirac points at $C_N$-symmetric $k$-points.
Since we have $\sum_n \nu_{a_n}=\t{tr } \Gamma=0$, 
the number of independent indices are $N-1$.
Thus the topological charges assigned to the Dirac point are
\begin{align}
(\nu_W,\nu_{a_1},\ldots,\nu_{a_{N-1}}) \in \Z^N.
\label{eq: topological charge}
\end{align}
The Dirac points with non-trivial topological charges are stable against
perturbations preserving chiral and rotation symmetry.
In Sec.~\ref{sec_c3} and Sec.~\ref{sec_c2}, 
we show examples of the Dirac points
protected by non-trivial indices $\nu_{a_i}$,
while the winding number $\nu_W$ is trivial.
In this sense, these are canonical examples of gapless points 
whose stability is not captured only by local symmetry (chiral symmetry),
but originates from spatial symmetry.

For two-fold rotation $C_2$,
we can also use the K-theory and the Clifford algebra to classify gapless
points:\cite{horava05,kitaev09,teo-kane10,morimoto2013topological,zhao-wang13,shiozaki14,kobayashi14,morimoto-weyl14}
%because all the symmetries are of order-two 
%(where two times of operation is proportional to identity.)
In this case, the symmetry operators $C_2$ and $\Gamma$ can be
considered as an element of a complex Clifford algebra $Cl_n=\{e_1,\dots, e_n\}$
with generators $e_1,\ldots,e_n$ satisfying
the anticommutation relation
\begin{align}
\{e_i,e_j\}=2\delta_{ij}.
\end{align}
Hence, the powerful representation
theory of the Clifford algebra is available in the classification.
%
%Together with gamma matrices of Dirac Hamiltonian, $C_2$ and $\Gamma$
%form a complex Clifford algebra $Cl_n$ %
Below, we show that the approach with Clifford algebra provides
the same topological charges in Eq.~(\ref{eq: topological charge}).

Consider a general Hamiltonian of 2D Dirac point,
\begin{align}
H=k_x\gamma_x+k_y\gamma_y,
\label{eq:clifford2D}
\end{align}
where $\gamma_i$'s are gamma matrices.
The symmetries $C_2$ and $\Gamma$ imply 
\begin{eqnarray}
\{\Gamma, \gamma_{i=x,y}\}=0,
\quad 
\{C_2, \gamma_{i=x,y}\}=0,
\quad
[\Gamma,C_2] = 0,\quad
\end{eqnarray}
so they form the complex Clifford algebra 
\begin{align}
Cl_3 \tensor Cl_1=
\{\gamma_x,\gamma_y,\Gamma\}
\tensor
\{\gamma_x\gamma_y C_2\},
\label{eq:Clifford1}
\end{align}
as we mentioned above.
Then if the Dirac point is unstable, 
there exists a Dirac mass term $m\gamma_0$ consistent with the symmetries,
\begin{eqnarray}
\{\Gamma, \gamma_0\}=0,
\quad 
[C_2, \gamma_0]=0,
\quad
\{\gamma_{i=x,y}, \gamma_0\}=0
\end{eqnarray} 
which modifies the Clifford algebra in
Eq.(\ref{eq:Clifford1}) as
\begin{align}
Cl_4 \tensor Cl_1=
\{\gamma_0, \gamma_x,\gamma_y,\Gamma\}
\tensor
\{\gamma_x\gamma_y C_2\}.
\label{eq:Clifford2}
\end{align}
The modified algebra implies that the mass term $\gamma_0$ behaves like 
an additional chiral operator $\Gamma'$ that anticommutes with $\Gamma$.
On the other hand, if the Dirac point is stable,
no such an additional chiral operator exists.
Therefore, the stability problem of the Dirac point reduces to the
existence problem of an additional chiral operator.\cite{morimoto2013topological,morimoto-weyl14}.

The latter problem is solved as follows.
By imposing chiral symmetry $\Gamma$ on other generators, 
we have an extension of Clifford algebra
\begin{align}
Cl_2 \tensor Cl_1&=\{\gamma_x,\gamma_y\}\tensor\{\gamma_x\gamma_y C_2\} \n
\to Cl_3 \tensor Cl_1&=\{\gamma_x,\gamma_y,\Gamma\}\tensor\{\gamma_x\gamma_y C_2\},
\label{eq: classification of Gamma (2D)}
\end{align}
which defines the classifying space ${\cal C}_0\times {\cal C}_0$ in the
K-theory. [${\cal C}_0=\cup_{m,n} U(m+n)/(U(m)\times U(n))$;
for details, see Refs.~\onlinecite{morimoto2013topological,morimoto-weyl14}].
Because the classifying space consists of all possible matrix representations of $\Gamma$ with other generators' fixed,
the zero-th homotopy group of the classifying space
\begin{align}
\pi_0({\cal C}_0 \times {\cal C}_0)=\Z^2,
\label{eq_Z^2}
\end{align}
measures topologically different chiral operators, 
specifying possible values for the topological number of $\Gamma$.
Now we can show that 
if there is an additional chiral operator
$\Gamma'$, then the topological number of $\Gamma$ must be zero:
Indeed, using $\Gamma'$, one can introduce the chiral operator
$\Gamma(t)=\Gamma\cos t+\Gamma'\sin t$ connecting   
$\Gamma=\Gamma(0)$ and $-\Gamma=\Gamma(\pi)$ continuously, which implies 
that $\Gamma$ must be topologically trivial
since 
topological numbers defined for chiral operators 
take opposite values for $\Gamma$ and $-\Gamma$
as we will see in an explicit way later [Eq.~(\ref{eq: index by trace})].
Taking the contrapositive, 
we can also say that if the topological number of $\Gamma$ is nontrivial,
then no additional chiral operator exists.
The last statement implies that the Dirac point is stable if the
topological number of $\Gamma$ is nontrivial.
%
%The above arguments clarify that the topological number in
%Eq.(\ref{eq_Z^2}) protects the Dirac point in Eq.(\ref{eq:clifford2D}). 
%This means that 
In other words, we can conclude that the topological charge protecting
the Dirac point in Eq.(\ref{eq:clifford2D}) is given by Eq.(\ref{eq_Z^2}), 
which coincides with Eq.~(\ref{eq: topological charge}) with $N=2$.

%Using the representation theory of Clifford algebra, we can solve the
%extension problem.  

%The reason is that,
%if topological number characterizing $\Gamma$ is trivial (non-trivial),
%then we can always find (cannot find) another mass term, 
%that can be adopted as $\gamma_0$.
%The classification of $\Gamma$ is given by a zero-th homotopy group of 
%a space of possible matrix representations of $\Gamma$ with fixing
%representation for other generators, and this 
%is a classifying space $C_0 \times C_0$.

%For details, see Ref.~\onlinecite{morimoto2013topological,morimoto-weyl14}.

The algebraic argument above can be intuitively 
understood by considering the specific Hamiltonian.
Let us take the effective Hamiltonian of 2D half-flux square lattice, 
$H=k_x\sigma_x\tau_z +k_y\sigma_y$
(i.e., $\gamma_x = \sigma_x \tau_z$, $\gamma_y = \sigma_y$)
with the two-fold rotation symmetry $C_2 = \sigma_z$,
and consider a possible generator $\Gamma$
to form an algebra
$Cl_3 \tensor Cl_1=\{\gamma_x,\gamma_y,\Gamma\}\tensor\{\gamma_x\gamma_y C_2\}$.
Since $\Gamma$ anticommutes with
$\gamma_x$ and $\gamma_y$ while 
commutes with $\gamma_x\gamma_y C_2 = i \tau_z$, 
it should be written as
\begin{equation}
 \Gamma = 
\begin{pmatrix}
s \sigma_z   &  0 \\
0 & s' \sigma_z  
\end{pmatrix},
\end{equation}
where the first and the second blocks
correspond to $\tau_z= \pm 1$, respectively,
and  $s, s' = \pm 1$.
Since $\tau_z= \pm 1$ are decoupled,
the sectors having different $(s,s')$
cannot be connected by a continuous transformation,
and thus they are topologically all distinct. 

If we generally consider the matrix $\tau_z$
with larger dimension
such as $\tau_z = {\rm diag}(1,1,\cdots,-1,-1,\cdots)$,
the possible expression for $\Gamma$ is 
\begin{eqnarray}
&& \Gamma = 
\begin{pmatrix}
\sigma_z \tensor A   &  0 \\
0 & \sigma_z \tensor A'
\end{pmatrix},
\end{eqnarray}
where the first and the second blocks in $\Gamma$
correspond to $\tau_z= \pm 1$, respectively.
Since we have $\Gamma^2=1$,
eigenvalues of $A$ and $A'$ are either $+1$ or $-1$.
The topologically distinct phases are labeled by two integers
\begin{align}
(s,s')=\left(\t{tr}A , \t{tr}A'  \right),
\label{eq: index by trace}
\end{align}
and this is $\Z^2$ in Eq.\ (\ref{eq_Z^2}).
The winding number is given by $\nu_W= s-s'$,
and the topological index of $C_2 = +1$ sector
(i.e., the difference between the numbers of 
the bases belonging to $\Gamma = +1$ and $-1$ in the $C_2 = +1$ sector)
is $\nu_{\rm even} = s + s'$.
So the space spanned by $(s,s')$ is equivalent to that by 
$(\nu_W, \nu_{\rm even})$.

\subsection{class AIII with $C_3R_z$ in 3D}

We study the chiral symmetric Dirac points with $C_3R_z$ symmetry (a
combination of a 3-fold rotation in $xy$-plane
and a reflection along $z$-axis) in 3D Brillouin zone, 
for which we have discussed an example in the stacked honeycomb lattice model
in Sec.~\ref{sec_3d_dirac}.
We write $g=C_3R_z$ and 
assume the commutation relation $[g,\Gamma]=0$.

We consider a Dirac point located at the $C_3R_z$ symmetric point,
and assume that the energy band is gapped in the vicinity of 
the Dirac point, except for the Dirac point itself.
At the Dirac point,
we can define the six topological numbers
$\nu_{\pm 1}, \nu_{\pm\omega}, \nu_{\pm\omega^2}$
as we have seen in Sec.~\ref{sec_3d_dirac},
but they are not completely independent.
Since $g^4 = C_3$ and $g^3 = R_z$,
the $C_3R_z$ symmetry is always accompanied by 
the individual symmetries $C_3$ and $R_z$. 
All the points on $k_z$ axis are fixed in $C_3$,
and in order to have a band gap at these momenta (except for the Dirac point),
all the indices for sectors $C_3=1,\omega,\omega^2$ should be zero;
\begin{align}
\nu_1+\nu_{-1}=\nu_\omega+\nu_{-\omega}=\nu_{\omega^2}+\nu_{-\omega^2}=0.
\label{eq: condition for indices for C3Rz}
\end{align}
Here note that sectors $g=\pm 1, \pm\omega,\pm\omega^2 $
belong to those $C_3=1,\omega, \omega^2$, respectively. 
%If this does not hold, we have a line node on $k_z$ axis.
Similarly, since the $k_x$-$k_y$ plane is fixed in $R_z$,
we have the requirement 
\begin{align}
\nu_1+\nu_\omega+\nu_{\omega^2}=\nu_{-1}+\nu_{-\omega}+\nu_{-\omega^2}=0,
\label{eq: condition 2 for indices for C3Rz}
\end{align}
in order to avoid a gap closing plane.
Due to these constraints, we are left with only two
independent indices, for example, $\nu_1,\nu_\omega$.
We can also define a winding number on the $R_z$ symmetric plane. 
Let us perform the block diagonalization with respect to $R_z=\pm 1$ 
on the $R_z$ symmetric plane. Then, the $R_z=+1$ sector is viewed 
as a 2D system class AIII+$C_3$,
and we can define a winding number $\nu_{W+}$ [Eq.~(\ref{eq: winding number})]
for $S^1$ surrounding the Dirac point. 
Similarly, we can also define  $\nu_{W-}$ for the $R_z=-1$ sector.
However, the total winding number $\nu_W=\nu_{W+}+\nu_{W-}$ 
should vanish because a circle $S^1$ defining the total winding number
can be freely deformed in the 3D space so 
it is contractible without touching the Dirac point.
Consequently, independent topological charges assigned to the Dirac point
in the present case are a set of a winding number $\nu_{W+}$ 
and two topological indices $\nu_1, \nu_\omega$:
\begin{align}
(\nu_{W+},\nu_1,\nu_\omega) \in \Z^3.
\end{align}

For the Dirac point at $K$ in the stacked honeycomb
lattice model in Sec.~\ref{sec_3d_dirac}, Eq.~(\ref{eq: charges C3Rz}) leads to 
$(\nu_{\pm 1},\nu_{\pm \omega}, \nu_{\pm \omega^2})=(\pm 1,\mp 1,0)$, which
%$(\nu_{\pm 1},\\nu_{\pm \omega}, \nu_{\pm \omega^2})=(\pm 1,0,\mp 1)$, which
 is consistent with the
constraints Eqs.~(\ref{eq: condition for indices for C3Rz})
and (\ref{eq: condition 2 for indices for C3Rz}). 
The winding numbers $\nu_{W\pm}$
can be evaluated 
using the effective Hamiltonian Eq.(\ref{eq_H_3d_dirac})
as follows.
On the $R_z$-symmetric plane $(k_z=\pi/(2c))$, the Hamiltonian is expressed as, 
\begin{eqnarray}
H=k_x\sigma_x\tau_z+k_y\sigma_y, 
\end{eqnarray}
with $R_z = \rho_z$ and $\Gamma=\sigma_z\rho_z$.
It takes the same form both in the $R_z =\pm 1$ sectors, 
but the chiral operator has an opposite sign,
i.e. $\Gamma=\pm \sigma_z$, leading to $\nu_{W\pm}=\pm 1$ 
for $K$-point ($\tau_z=+1$).
Since $\nu_{W\pm}$ is non-zero,
non-trivial indices $\nu_{a_i}$ are not necessary
for the topological protection of the Dirac point in this particular example.
However, if we consider a $C_3R_z$ symmetric superlattice where $K$ and
$K'$-points are folded onto the same $\Gamma$ point,
as in the case of the 2D honeycomb lattice in Sec.~\ref{sec_c3},
the winding number around the Dirac point becomes zero while other indices
$\nu_{a_i}$ are still non-zero.
There, the gaplessness at the Dirac point is solely guaranteed by non-trivial indices $\nu_{a_i}$.
%[Remark:
%Alternatively,
%this model (stacked honeycomb lattice model
%in Sec.~\ref{sec_3d_dirac}) 
%would be understood from topological charges for class A+$C_3$.
%We can construct a natural suspension from class AIII+$C_3$ of $S^1$ to
%class A+$C_3$ of $S^2$.
%Then topological indices for $S^1$ is mapped to a number of valence bands at north and south poles($C_3$ invariant momenta) of each sector of $C_3$.
%This indicates that chiral symmetry is not a necessary ingredient for topological protection of our example.
%But this would be a departure form our discussions focusing on chiral symmetry.
%Besides, topological indices of $C_3$ sectors can be more easily computed.
%]

\subsection{class AIII with $C_2R_z$ in 3D}

Finally, we study the chiral symmetric Dirac points with the
inversion symmetry $P=C_2R_z$ in 3D.
Here we consider two different cases,
(i) where we have $C_2$ and $R_z$ symmetries individually,
and  (ii) where we only have $P$ but not $C_2$ or $R_z$.

First we consider the case (i).
The half-flux cubic lattice model argued in Sec.~\ref{sec_3d_dirac}
belongs to this case.
We assume $[C_2,R_z]=[C_2,\Gamma]=[R_z,\Gamma]=0$.
At the inversion symmetric point,
we can define the four topological indices
$\nu_{++},\nu_{+-},\nu_{-+},\nu_{--}$
for the sectors labeled by the eigenvalues of $(C_2,R_z)$.
To avoid the band gap closing on the $C_2$ symmetric axis,
\begin{align}
\nu_{++} + \nu_{+-} = \nu_{-+} + \nu_{--} = 0.
\end{align}
To gap out $R_z$ symmetric plane, similarly, we require
\begin{align}
\nu_{++} + \nu_{-+} = \nu_{+-} + \nu_{--} = 0.
\end{align}
Therefore $(\nu_{++},\nu_{+-},\nu_{-+},\nu_{--})$
is expressed by a single integer $s$ as $(s,-s,-s,s)$.
In Sec.~\ref{sec_3d_dirac}, we defined the topological indeces
$(\nu_{\rm even}, \nu_{\rm odd})$
for the sectors labeled by $P = C_2 R_z$,
and they are related to the present indices by
$\nu_{\rm even} = \nu_{++} + \nu_{--} =2s$
and $\nu_{\rm odd} = \nu_{+-} + \nu_{-+} =-2s$.

Similarly to the $C_3R_z$ case in the previous subsection,
we can define the winding numbers $\nu_{W\pm}$ 
for $R_z=\pm 1$ sector, respectively.
The total winding number $\nu_W=\nu_{W+}+\nu_{W-}$ 
vanishes again because of the same reason.
Therefore, independent topological charges assigned to a Dirac point are
\begin{align}
(\nu_{W+},\nu_{++}) \in \Z^2.
\label{eq:Z2inversion}
\end{align}

In the case (ii), we can define the topological indices
$\nu_{\rm even},\nu_{\rm odd}$
for the sectors labeled by the eigenvalues of the inversion $P$
(where $[P,\Gamma]$ is assumed).
The summation $\nu_{\rm even}+\nu_{\rm odd}=\t{tr}\,\Gamma$ 
should vanish otherwise
the band gap closes everywhere in $k$-space.
Unlike the case (i), we do not have the winding numbers $\nu_{W\pm}$
since $R_z$ symmetry is absent and thus 
we do not have a 2D subspace invariant under the symmetry operation. 
As a result, the Dirac point is characterized only by
a single topological number,
\begin{align}
\nu_{\rm even} \in \Z.
\label{eq:Zinversion}
\end{align}

Because $C_2$ and $R_z$ are both order-two operators,
we can also derive the same conclusion from
the analysis using the Clifford algebra.
Let us consider a 3D Dirac point
\begin{align}
H=k_x\gamma_x+k_y\gamma_y+k_z\gamma_z,
\end{align}
and explore whether a mass term $m\gamma_0$ is allowed or not by imposed
symmetries.

In the case (i), we have three symmetries: 
chiral symmetry $\Gamma$,
two-fold rotation in $xy$-plane $C_2$,
reflection symmetry along $z$-direction $R_z$.
The symmetry operators satisfy the following algebraic relations
with the gamma matrices,
\begin{align}
\{\gamma_{i=0,x,y,z},\Gamma\}&=0, 
\n
[\gamma_{i=0,z},C_2]=\{\gamma_{i=x,y},C_2\}&=0, 
\n
[\gamma_{i=0,x,y},R_z]=\{\gamma_{z},R_z\}&=0, 
\end{align}
with the commutation relations with each other
\begin{align}
[R_z,C_2]=[C_2,\Gamma]=[R_z,\Gamma]&=0.
\end{align}
Then we can construct a Clifford algebra from these relations as
\begin{align}
Cl_{6}\tensor Cl_1&=\{\gamma_0,\gamma_x,\gamma_y,\gamma_z,\gamma_z R_z, \Gamma \}
\tensor \{\gamma_x\gamma_y C_2\}.
\end{align}
%which describes all the symmetry constraints of the 3D $\pi$-flux model.
In a similar way as Sec.~\ref{sec:charges}A,
the mass term $\gamma_0$ can be considered as an additional chiral
operator $\Gamma'$, so if the topological number of $\Gamma$ is nonzero,
then the Dirac point is stable.
From an extension of Clifford algebra which is obtained by adding $\Gamma$ to
other generators,
%The topological number of $\Gamma$ is derived 
%the type of topological charge, i.e., the existence condition of $\gamma_0$
% is obtained from 
%classification of (for example) $\Gamma$ in the algebra without $\gamma_0$ as
\begin{align}
Cl_{4}\tensor Cl_1 &=\{\gamma_x,\gamma_y,\gamma_z,\gamma_z R_z \}
\tensor \{\gamma_x\gamma_y C_2\} \n
\to Cl_{5}\tensor Cl_1 &=\{\gamma_x,\gamma_y,\gamma_z,\gamma_z R_z, \Gamma \}
\tensor \{\gamma_x\gamma_y C_2\}, 
\end{align}
we identify the relevant classifying space as 
${\cal C}_0 \times {\cal C}_0$, then 
the relevant topological number is evaluated as the zero-th homotopy,
\begin{align}
\pi_0({\cal C}_0 \times {\cal C}_0)=\mathbb{Z}^2,
\end{align}
which coincides with Eq.(\ref{eq:Z2inversion}).

In the case (ii),
the additional symmetry is only inversion $P=C_2R_z$.
The algebraic relations for $P$ read
\begin{align}
[\gamma_0,P]=\{\gamma_{i=x,y,z},P\} =0, \quad 
[P,\Gamma] =0, 
\end{align}
which form the Clifford algebra,
\begin{align}
Cl_6 &= \{\gamma_0,\gamma_x,\gamma_y,\gamma_z,
\gamma_x\gamma_y\gamma_z P, \Gamma \}.
\end{align}
The existence condition for the mass term $\gamma_0$
is obtained from the extension problem
%classification of $\Gamma$ in the extension,
\begin{align}
Cl_{4} &=\{\gamma_x,\gamma_y,\gamma_z,
\gamma_x\gamma_y\gamma_z P \}
\n
\to Cl_{5} &=\{\gamma_x,\gamma_y,\gamma_z,
\gamma_x\gamma_y\gamma_z P, \Gamma \},
\end{align}
which gives the classifying space as ${\cal C}_0$, 
and thus the topological charge protecting the Dirac point is given by
\begin{align}
\pi_0({\cal C}_0)=\mathbb{Z}.
\end{align}
This result reproduces Eq.(\ref{eq:Zinversion}).

\section{Conclusion}
\label{sec_conc}
In this paper, we show that the coexistence of chiral symmetry and the
spatial symmetry can stabilize zero energy modes, even when the chiral
symmetry alone does not ensure their stability. 
We present general arguments for the stability and we identify the
associated topological numbers.
The validity of our arguments are demonstrated for the Dirac points in two
dimensions with a variety of spatial symmetries. 
We also illustrate that Dirac semimetals in three dimensions are
possible in the presence of coexisting spatial symmetries.
In the last part, we list up and classify 
independent topological invariants 
associated with a given Dirac point.
We find that the set of topological numbers found here
gives a complete minimal set of quantum numbers 
allowed by the algebraic constraint
in the case of order two symmetries.

\section*{ACKNOWLEDGMENT}

The authors acknowledge
C. Hotta, K. Asano and K. Shiozaki for useful discussions.
This project has been
funded by JSPS Grant-in-Aid for Scientific Research No.
24740193, No. 25107005 (M.K.), No. 24840047 (T.M.), and No.22103005,
No. 25287085 (M.S.). 

%{\bf @@  add the funding sources}

\bibliography{chiral}

\end{document}